\documentclass[sigconf]{acmart}

\usepackage[linguistics]{forest}
\usepackage{tikz}
\usetikzlibrary{shapes.geometric, arrows}
\usepackage{pgfplots}
\usepgfplotslibrary{fillbetween}
\AtBeginDocument{%
  \providecommand\BibTeX{{%
    \normalfont B\kern-0.5em{\scshape i\kern-0.25em b}\kern-0.8em\TeX}}}

\usepackage{longtable}

\copyrightyear{2022}
\acmYear{2022}
\setcopyright{rightsretained}
\acmConference[AIES'22]{Proceedings of the 2022 AAAI/ACM Conference on AI, Ethics, and Society}{August 1--3, 2022}{Oxford, United Kingdom} \acmBooktitle{Proceedings of the 2022 AAAI/ACM Conference on AI, Ethics, and Society (AIES'22), August 1--3, 2022, Oxford, United Kingdom}\acmDOI{10.1145/3514094.3534176} \acmISBN{978-1-4503-9247-1/22/08}

\pgfplotsset{compat=1.17} 
\begin{document}
\fancyhead{}

%%
%% The "title" command has an optional parameter,
%% allowing the author to define a "short title" to be used in page headers.
\title[Measuring Gender Bias in Grammatically Gendered Language Embeddings]{Measuring Gender Bias in Word Embeddings of Gendered Languages Requires Disentangling Grammatical Gender Signals}

%%
%% The "author" command and its associated commands are used to define
%% the authors and their affiliations.
%% Of note is the shared affiliation of the first two authors, and the
%% "authornote" and "authornotemark" commands
%% used to denote shared contribution to the research.

\author{Shiva Omrani Sabbaghi}
\affiliation{%
  \institution{George Washington University}
    \institution{Department of Computer Science}
  \city{Washington}
 \state{D.C.}
  \country{USA}
 }
\email{somrani@gwu.edu}

\author{Aylin Caliskan}
\affiliation{%
  \institution{University of Washington}
      \institution{Information School}
  \city{Seattle}
  \state{WA}
  \country{USA}
}
\email{aylin@uw.edu}

%%
%% By default, the full list of authors will be used in the page
%% headers. Often, this list is too long, and will overlap
%% other information printed in the page headers. This command allows
%% the author to define a more concise list
%% of authors' names for this purpose.
\renewcommand{\shortauthors}{Omrani Sabbaghi and Caliskan}
%%
%% The abstract is a short summary of the work to be presented in the
%% article.
\begin{abstract}
  Does the grammatical gender of a language interfere when measuring the semantic gender information captured by its word embeddings? A number of anomalous gender bias measurements in the embeddings of gendered languages suggest this possibility. We demonstrate that word embeddings learn the association between a noun and its grammatical gender in grammatically gendered languages, which can skew social gender bias measurements. Consequently, word embedding post-processing methods are introduced to quantify, disentangle, and evaluate grammatical gender signals. The evaluation is performed on five gendered languages from the Germanic, Romance, and Slavic branches of the Indo-European language family. Our method reduces the strength of grammatical gender signals, which is measured in terms of effect size (Cohen's \textit{d}), by a significant average of \textit{d} = 1.3 for French, German, and Italian, and \textit{d}= 0.56 for Polish and Spanish. Once grammatical gender is disentangled, the association between over 90\% of 10,000 inanimate nouns and their assigned grammatical gender weakens, and cross-lingual bias results from the Word Embedding Association Test (WEAT) become more congruent with country-level implicit bias measurements. The results further suggest that disentangling grammatical gender signals from word embeddings may lead to improvement in semantic machine learning tasks.
\end{abstract}

%%
%% The code below is generated by the tool at http://dl.acm.org/ccs.cfm.
%% Please copy and paste the code instead of the example below.
%%



%%
%% Keywords. The author(s) should pick words that accurately describe
%% the work being presented. Separate the keywords with commas.

%% A "teaser" image appears between the author and affiliation
%% information and the body of the document, and typically spans the
%% page.
% \begin{teaserfigure}
%   \includegraphics[width=\textwidth]{sampleteaser}
%   \caption{Seattle Mariners at Spring Training, 2010.}
%   \Description{Enjoying the baseball game from the third-base
%   seats. Ichiro Suzuki preparing to bat.}
%   \label{fig:teaser}
% \end{teaserfigure}

%%
%% This command processes the author and affiliation and title
%% information and builds the first part of the formatted document.
\maketitle

\section{Introduction}
English word embeddings learn human-like biases including gender bias from word co-occurrence patterns \citep{10.5555/3157382.3157584, Caliskan183, GargE3635, caliskan2020social}, leading to associations such as man is to doctor as woman is to nurse. How can we measure these social gender biases in the embeddings of gendered languages where even inanimate nouns are assigned a gender? For instance, does the embedding for Spanish word \textit{fuerza} (strength), a stereotypically masculine trait, carry more similarity to masculinity despite being a grammatically feminine noun? Are social gender biases in embeddings intensified, subdued, or unchanged by the presence of grammatical gender in gendered languages?

Natural language processing (NLP) applications use static or dynamic word embeddings as general-purpose language representations for numerous tasks including machine translation \citep{TACL1081}, document ranking \citep{10.1145/2872518.2889361}, and sentiment classification \citep{NIPS2014_2cfd4560}. Moreover, NLP has applications in social contexts for consequential decision-making. Since potentially harmful biases learned by word embeddings may propagate to downstream applications such as resume screening, essay grading, or university admissions \citep{zhao-etal-2018-gender, osti_10098355, zhao-etal-2017-men}, answers to the questions outlined earlier have important implications. 

Accurate measurement of biases in word embeddings is critical not only because it helps with tracing biases in downstream applications but also because it helps with measuring biases in the text corpora which represent certain populations. Bias measurement is specially challenging in gendered languages, since the embeddings of nouns are impacted by grammatical gender \citep{gonen-etal-2019-grammatical-gender} which can impact the results. \citet{toney2020valnorm} note that gender bias measurements in Polish (a gendered language) using the Word Embedding Association Test (WEAT), a widely-used bias measurement method, indicate a men:humanities-women:science association which is incongruent with gender bias measurements reported by psychologists \citep{Nosek10593}. This bias test in Polish contains a set of words that represent science with grammatically feminine words that amplify women:science association. The resulting stereotype-incongruent measurement suggests that, among other factors, structural properties of a language such as grammatical gender may impact bias measurements in word embeddings. Thus, quantifying biases using WEAT or similar approaches without taking grammatical gender into account may lead to inaccurate results.

\begin{figure}
    \centering
    {\footnotesize
\begin{forest}
  [\textbf{Gender Associations in Word Embeddings}
    [\textbf{Grammatical Gender}
     [\textit{el sol} - the sun (masculine)\\
     vs. \\
     \textit{la luna} - the moon (feminine)
    ]
    ]
    [\textbf{Semantic Gender}
        [Social Gender Bias
            [man:doctor-woman:nurse
        ]
        ]
        [Gender Identity
            [rooster vs. hen  
            ]
        ]
        ]
    ]
\end{forest}
}
    \caption{Gender associations learned by word embeddings can be a mix of signals from semantic and grammatical gender. Semantic gender includes both gender identity and stereotypical gender associations while grammatical gender refers to gender assigned to words in gendered languages.}
    \label{fig:my_label}
\end{figure}
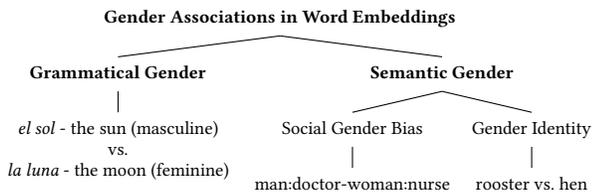

We refer to stereotypical gender biases (e.g. men:career\\-women:family) as \textit{social gender bias}, one's perceived/own sense of gender (e.g. rooster is a male chicken) as \textit{gender identity}, and the association between grammatically masculine/feminine nouns (e.g. \textit{fuerza}) with their syntactic gender as \textit{grammatical gender signal}. Note that both social gender bias and gender identity fall under the category of semantic gender, while grammatical gender signal is syntactic as shown in Figure~\ref{fig:my_label}.

In this paper, we ask the question: Is grammatical gender signal causing anomalous gender bias measurements? To answer this question, we present a grammatical gender disentanglement technique. Then we ask: Does disentangling grammatical gender from embeddings impact their semantic quality? Is grammatical gender effectively disentangled from word embeddings? By answering these three questions, we make the following main contributions:
\begin{enumerate}
    \item Proposing post-processing methods to disentangle grammatical gender signals iteratively from word embeddings via linear support vector classification (SVC) \citep{svccite}. The accuracy in detecting grammatical gender drops from an average of $96\%$ to $50\%$ of random guessing.
    \item The cross-linguistic generalization of WEAT for measuring bias in five grammatically gendered languages from the Germanic (German), Romance (French, Italian, Spanish), and Slavic (Polish) branches of the Indo-European language family. We show that effect size in Cohen's $d$, which quantifies the magnitude of semantic social biases via WEAT, increases by an average of $d= 0.24$ once the grammatical gender signal is disentangled.
    \item Applying a variation of WEAT called the Grammatical Gender Word Embedding Association Test (GG-WEAT) for evaluating the presence and strength of grammatical gender signals in the embedding space. GG-WEAT shows that grammatical gender signal’s magnitude is reduced by an average of $d= 1.00$ for the languages studied.
\end{enumerate}
Although in this work we focus on binary gender, we acknowledge the importance of studying other representations of semantic and grammatical gender which we leave to future work. Code and data can be found at  \url{https://github.com/shivaomrani/GG_Disentangling}. 

\section{Background}

\textbf{Grammatical gender in a language} refers to the assignment of gender to a word, even if there is no clear association between the word and its grammatical gender. While the grammatical gender of animate nouns usually match their gender identity \citep{RefWorks:doc:600c76508f08052520142198, KRAMER2014102}, many researchers have suggested that grammatical gender assignment to inanimate nouns appears to be arbitrary \citep{12b5294fa99747b0b07bf7b76328a7df}. For example, the Spanish word for moon, an inanimate noun, is the feminine word \textit{luna} while the German translation is the masculine noun \textit{Mond} \citep{mccurdy2020grammatical}. Such inconsistencies across different languages suggest that grammatical gender assignment seems to be arbitrary. Grammatical gender assignment then dictates how a word should appear in a sentence. For example, in Spanish, \textit{luna} appears in feminine context, co-occurs with the feminine article \textit{la} and gets surrounded by other parts of speech that must agree with \textit{luna}'s gender.

According to the distributional hypothesis of language, words co-occurring in similar context windows tend to have similar meanings \citep{doi:10.1080/00437956.1954.11659520, Firth1957, https://doi.org/10.1002/(SICI)1097-4571(199009)41:6<391::AID-ASI1>3.0.CO;2-9}. Word embeddings are trained on word co-occurrence statistics that not only capture semantics but also syntactic associations, including the association of words and their grammatical gender \citep{gonen-etal-2019-grammatical-gender}. The entanglement of semantic gender with grammatical gender can skew social gender bias measurements in word embeddings \citep{toney2020valnorm, mccurdy2020grammatical}.

\textbf{Quantifying bias in word embeddings} can be accomplished by using WEAT\citep{Caliskan183}, which measures social stereotypes as well as widely shared non-social group associations in word embeddings. WEAT is an adaptation of the Implicit Association Test (IAT) \cite{pmid9654756} from social psychology which measures implicit biases in human cognition. In IAT, participants are presented with two sets of target concepts (e.g. career and family) and two sets of attributes (e.g. men and women) via access words (stimuli that represent these concepts, such as \textit{salary} and \textit{business} for career; and \textit{mother} and \textit{girl} for women). Test participants are asked to pair the concepts with attributes as fast as they can. The reaction time paradigm in this categorical pairing task quantifies implicit biases of human subjects. Participants are faster in pairing words which they find similar (e.g. father:salary-mother:house). Therefore, the differential latency between pairing stereotype-congruent (e.g. men:career-women:family) and stereotype-incongruent (e.g. men:family-women:career) targets and attributes quantifies the effect size (Cohen's $d$) of implicit biases. Similarly, WEAT quantifies the differential association between two target
sets and two attribute sets in word embeddings. The works of \citet{lauscher-glavas-2019-consistently} and \citet{PMID:32747806} are among studies that have used WEAT-based methods to measure biases in other languages, including gendered languages.
\begin{table*}[t!]
\begin{tabular}{lccccc}
\hline {\textbf{Task}} & {\citet{gonen-etal-2019-grammatical-gender}} & {\citet{mccurdy2020grammatical}} & {\citet{zhou-etal-2019-examining}} 
& {This work}\\ \midrule
{Measuring gender bias considering GG} & {--} & {\checkmark} & {\checkmark} & {\checkmark}\\
{Identifying GG} & {--} & {--} & {\checkmark} & \checkmark\\
{Quantifying Strength of GG} & {--} & {\checkmark} & {--} & \checkmark\\
{Disentangling GG from embeddings} & {\checkmark} & {\checkmark} & -- & $\star$\\
{Evaluating impact of disentangling GG on semantics} & {\checkmark} & {--} & {--} & \checkmark\\

\hline
\end{tabular}
    \caption{\label{tab:prior} Comparison to prior directly related work. GG refers to Grammatical Gender. Our methodology is not directly comparable to \citet{zhou-etal-2019-examining} since grammatical gender is not disentangled from word embeddings in their study. $\star$ denotes post-processing GG disentanglement method as opposed to pre-processing methods proposed by prior work.}
\end{table*}

\section{Related Work}

A number of previous studies have paid attention to grammatical gender in word embeddings. \citet{10.1145/3461702.3462530} and \citet{zhao-etal-2020-gender} measure gender bias in multilingual embeddings (including gendered languages) by using occupation terms, and use paired occupations (e.g. waiter and waitress) in each language to account for grammatical gender.

Recognizing that grammatical gender forces the embeddings of same-gender inanimate nouns to be closer to each other than nouns with different gender in the vector space, \citet{gonen-etal-2019-grammatical-gender} disentangle grammatical gender signals with training data preprocessing methods such as lemmatizing and changing the gender of all context words to the same gender with the help of morphological analyzers. Despite positive results for German and Italian, it is emphasized that disentangling grammatical gender signals through pre-processing methods is not a trivial task as it relies on careful usage of language-specific morphological analyzers.

\citet{mccurdy2020grammatical} explore gender bias in Spanish, Dutch, and German, concluding that grammatical gender signals outweigh social gender bias in cross-linguistic word embeddings. They suggest that basic lemmatization of training corpora can help with reducing both grammatical gender signals and social gender bias in the resulting word embeddings. Our post-processing method is different from these pre-processing approaches because we disentangle grammatical gender from pre-trained embeddings.

\citet{zhou-etal-2019-examining} identify (but do not disentangle) grammatical gender signals by applying Linear Discriminant Analysis (LDA) \citep{https://doi.org/10.1111/j.1469-1809.1936.tb02137.x} on 6,000 grammatically feminine and masculine nouns. The resulting vector is projected out of an "overall gender" vector to obtain semantic gender. Since inanimate nouns should not carry semantic gender information, semantic gender component is removed from these nouns while grammatical gender is preserved.

Our study is similar to \citep{zhou-etal-2019-examining} in the sense that we also use LDA-like methods (SVC) to identify grammatical gender. However, instead of preserving grammatical gender and using it for other purposes, we disentangle it from embeddings. Furthermore, we notice that one iteration of applying LDA/SVC is not sufficient for identifying grammatical gender, so we apply it iteratively to fully capture and disentangle grammatical gender information. Therefore, although there are similarities between our work and \citep{zhou-etal-2019-examining}, our methodologies are not directly comparable since the objectives are different. Still, we highlight some results after one iteration of disentangling grammatical gender to emphasize the necessity of our iterative process.

Table~\ref{tab:prior} shows comparison of our work to the directly related prior studies. Note that although our methodology is theoretically comparable to \citep{gonen-etal-2019-grammatical-gender} and \citep{mccurdy2020grammatical} (because these studies also disentangle grammatical gender), in practice we cannot compare results because these studies train their own embeddings while we disentangle grammatical gender from pre-trained embeddings. Due to the difference in scope, quality and availability of embeddings\footnote{Out of the languages we studied, only corresponding Italian and German embeddings from \citep{gonen-etal-2019-grammatical-gender} and German and Spanish embeddings from \citep{mccurdy2020grammatical} are available.} trained in these two studies, our results are incomparable.

\begin{figure*}[t!]
    \centering
    \includegraphics[scale=0.4]{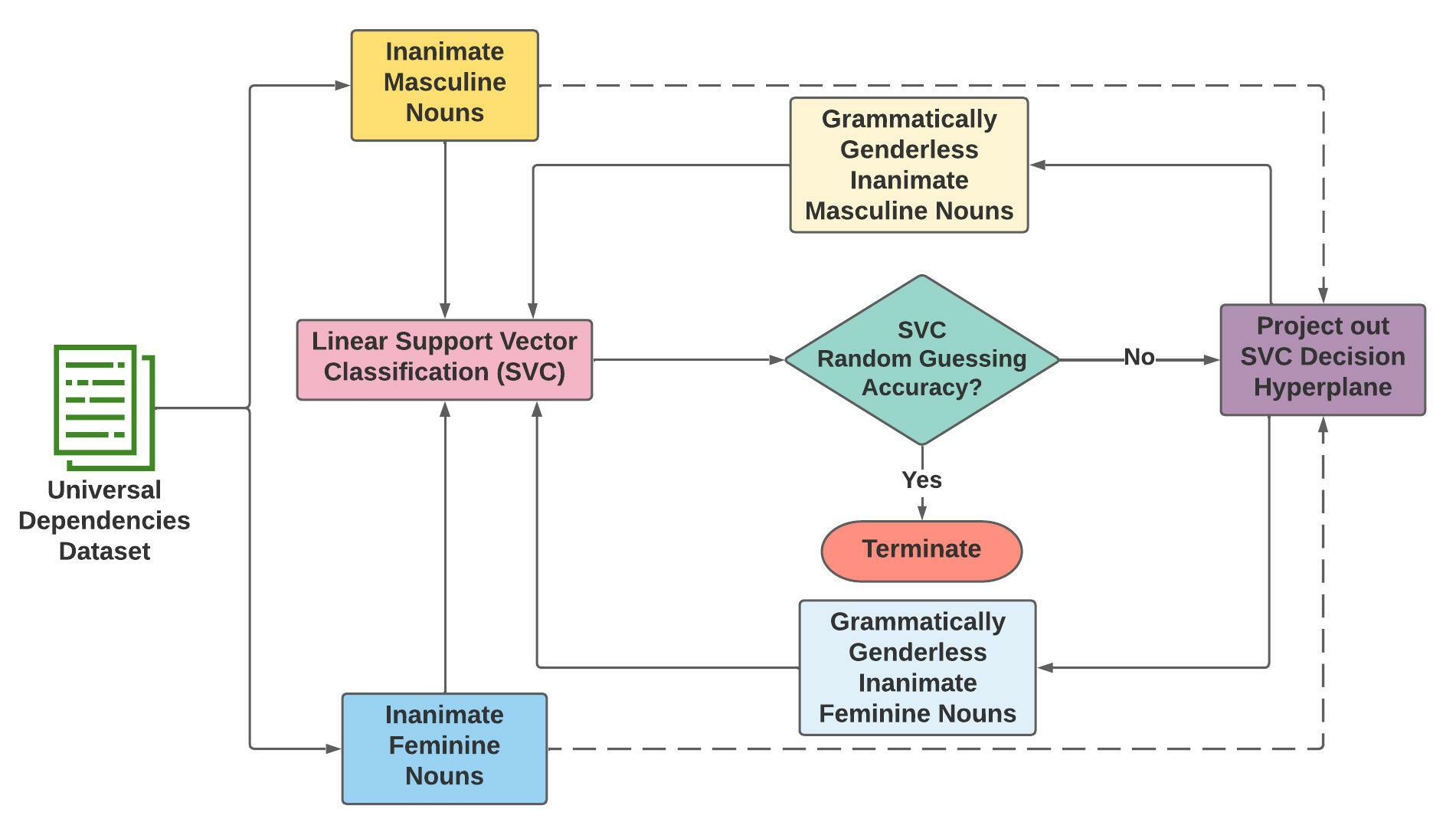}
    \caption{The layout of our approach for identifying and disentangling grammatical gender from word embeddings. Solid lines represent main paths in the flowchart while dotted lines represent dataset usage.}
    \label{fig:flow}
\end{figure*}

\section{Data}
We select FastText \citep{grave2018learning, bojanowski2017enriching} pre-trained embeddings as these are the only available cross-lingual embeddings that were trained on large-scale corpora under similar conditions.\footnote{FastText Embeddings for all languages are trained on \textit{Wikipedia} and \textit{Common Crawl}. The differences in training corpora across languages are negligible as embeddings are all of high quality according to intrinsic evaluation criteria.} The following languages with various degrees of grammatical gender from three different branches of Indo-European language family are selected based on the availability of high quality embeddings as well as IAT results and other evaluation datasets:

\noindent\textbf{Germanic}: English (EN) and German (DE) are from the West Germanic language branch. While English uses gendered pronouns (he/she/it) and has no concept of gender assignment to inanimate nouns, German uses a three-gender grammatical gender system with feminine, masculine, and neuter gender classes \citep{RefWorks:doc:600c95578f082cc239799788}.

\noindent\textbf{Romance}: French (FR), Italian (IT), and Spanish (ES) are from the Romance language branch, and use the two classes of masculine and feminine grammatical gender. \citep{RefWorks:doc:600c9b3c8f08a3d6ada01530, maiden2014reference, doi:10.1162/0898929052880101}.

\noindent\textbf{Slavic}: Polish (PL) is from the Slavic branch and uses three main gender classes of masculine, feminine, and neuter. However, in plural form there can be masculine-personal and non-masculine-personal genders. Thus, Polish grammatical gender can also be characterized as having "three singular and two plural grammatical genders" \citep{10.3389/fpsyg.2019.02208}. As an inflected language with a lot of grammatical gender markers \citep{10.3389/fpsyg.2019.02208}, Polish syntax is considered to be complex \citep{RefWorks:doc:600c961e8f08ddf0efbdd125}.

\subsection{Grammatical Gender Classification Data}
Grammatically feminine and masculine inanimate nouns are needed for performing grammatical gender classification and finding the grammatical gender subspace. We used the training portion of the Universal Dependencies (UD) dataset \footnote{\url{https://universaldependencies.org} } to find grammatically feminine and masculine nouns, since grammatical gender is an annotated feature in this dataset (the annotations are either manual, or automatically converted from manual for all the datasets that we used). We used the following datasets: \textit{AnCora} \citep{taule-etal-2008-ancora} for Spanish, \textit{GSD}\citep{guillaume:hal-02267418} for French, \textit{HDT}\citep{borges-volker-etal-2019-hdt} for German, \textit{ISDT}\citep{bosco-etal-2013-converting} for Italian and \textit{PDB}\citep{plud} for Polish. To ensure that the subspace we find corresponds to only grammatical gender and contains minimal semantic gender information, all animate nouns from the data must be removed. However, except for Polish, animacy tags were not available for nouns in the datasets. Thus, we removed any word that appeared in the animate noun list compiled by \citet{zmigrod-etal-2019-counterfactual} from our dataset. This list was only available for French and Spanish, so we used Google translate to create the list for other languages. As an additional step to identify and exclude animate nouns, we used Open Multilingual WordNet \citep{Sagot:Fiser:2008, Gonzalez-Agirre:Laparra:Rigau:2012, Toral:Bracale:Monachini:Soria:2010, Piasecki:Szpakowicz:Broda:2009, MAPPING2012,plwordnet2}. A noun was considered to be animate if \textit{person} was a hypernym of the noun in WordNet. \citep{zmigrod-etal-2019-counterfactual}. We applied this additional step to all languages except for German, because Open Multilingual Wordnet\footnote{\url{http://compling.hss.ntu.edu.sg/omw/}} was not available for German.

\subsection{WEAT Stimuli}
As mentioned earlier, we use WEAT to measure biases associated with different concepts in embeddings. Like IAT, WEAT requires use of stimuli (word lists) to represent concepts. In this study, we need cross-lingual stimuli for the following $7$ pairs of concepts: (1) science-humanities, (2) men-women words (e.g. mother, girl), (3) career-family, (4) list of common names for boys-girls (5) flowers-insects, (6) musical instruments-weapons, (7) list of pleasant-unpleasant words.

Science-humanities and men-women stimuli for all languages are available in Project Implicit Website\footnote{\url{https://implicit.harvard.edu/implicit/}} where people can take the The Implicit Association Test in different languages. However, career-family stimuli are only available for English and German. For flowers-insects, musical instruments-weapons and pleasant-unpleasant we used stimuli provided by \citet{toney2020valnorm} for English, German, Spanish and Polish. Whenever a stimuli was missing for a language, we built a custom list by translating the corresponding English stimuli. All translations are done using Google Translate, and stimuli are case-sensitive. 

Furthermore, we used a list of common names for boys and girls that were available for English and German on the Project Implicit Website. For other languages, we searched Google to find most frequent names for boys and girls in countries where the languages were primarily spoken. For more details regarding the creation of the stimuli, including full list of stimuli used in social gender bias experiments, please refer to the appendix. 

In addition to the data mentioned above, we use other datasets such as cross lingual analogy \citep{grave2018learning, koper-etal-2015-multilingual, berardi2015word,cardellinoSBWCE}, and cross lingual affective norms \citep{toney2020valnorm,PMID:24150921,PMID:24366716}, and cross lingual Simlex-999 \citep{10.1162/COLI_a_00237, barzegar-etal-2018-semr, MYKOWIECKA18.687} as explained in sections \ref{sec:gg-eval} and \ref{sec:semqual}. As these datasets were available for all languages in our study, there was no need for translation or customization choices.

\section{Approach}
\label{sec:approach}
We hypothesize that word embeddings learn grammatical gender in addition to semantic gender information. Thus, we first \textbf{identify} grammatical gender signals in the embedding space if they exist. Furthermore, we hypothesize that grammatical gender interferes with semantic gender bias measurements. Thus, we attempt to \textbf{disentangle} grammatical gender signals from embeddings and use \textbf{WEAT} to measure embedding bias before and after disentangling grammatical gender signals to see how the measurements change. This section details methodologies for identifying and disentangling grammatical gender as well as measuring biases. Figure~\ref{fig:flow} summarizes grammatical gender identification and disentanglement steps.

\subsection{Grammatical Gender Signal Identification}
\label{sec:isolation}

Our method of disentangling grammatical gender from embeddings is inspired by \citet{zhou-etal-2019-examining}'s use of LDA for identifying grammatical gender. We define the binary classification task of classifying inanimate grammatically feminine and masculine nouns to extract a decision hyperplane that corresponds to grammatical gender. The hyperplane is projected out of embeddings, and model is re-applied to evaluate the extent to which word embeddings unlearn grammatical gender via signal disentanglement. Ideally, we expect $100\%$ and $50\%$ (random-guessing) classification accuracy before and after disentangling grammatical gender. We experiment with both LDA and SVC but report the results for SVC only for being more efficient despite the models reaching very similar results.

SVC is a linear classifier that finds the maximum-margin hyperplane that separates classes. To identify the grammatical gender signal, SVC models are trained on 6,000 inanimate grammatically feminine and masculine nouns. The decision hyperplane $\vec{d_g}$ that results from classifying grammatically feminine and masculine nouns nontrivially corresponds to grammatical gender. The large size of the dataset, the randomness of chosen words, and the fact that all words are inanimate nouns mean that other properties of embeddings such as part of speech or semantic meaning are minimally captured by $\vec{d_g}$.

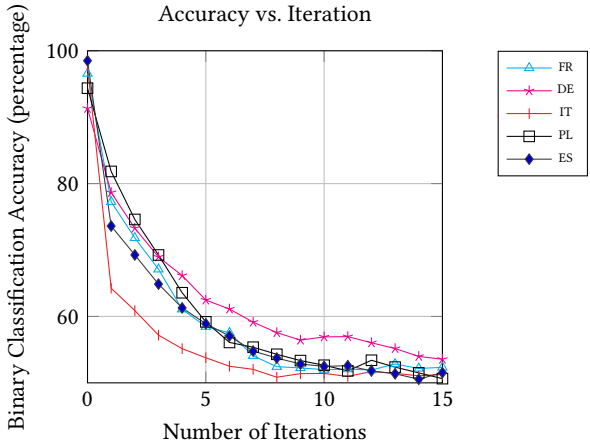
\begin{figure}
    \centering
    % \documentclass[border=10pt]{standalone}
% \usepackage{pgfplots}
% \usepgfplotslibrary{fillbetween}
\pgfplotsset{height=6cm, width=63mm}
% \begin{document}
\begin{tikzpicture}

\definecolor{clr1}{cmyk}{100,0,0,0}
\definecolor{clr2}{cmyk}{0,100,0,0}
\definecolor{clr3}{cmyk}{0,50,100,0}
\definecolor{clr4}{cmyk}{100,69,0,0}
\definecolor{clr5}{cmyk}{60,14,68,0}

  \begin{axis}[
  title= Accuracy vs. Iteration,
  xlabel={Number of Iterations},
  ylabel={Binary Classification Accuracy (percentage)},
  grid=major,
  legend entries={FR,DE,IT,PL,ES}, legend style={at={(1.41,1.0)}, font=\tiny},
  ymin=50,
  ymax=100,xmin=0,
  xmax=15,]
  
    \addplot+[mark=triangle, sharp plot, clr1] coordinates
    %french
      {
      (0,0.9658*100)
        (1,0.7725*100)
        (2,0.7181666*100)
        (3,0.671*100)
        (4,0.6108*100)
        (5,0.58466666*100)
        (6,0.575333333*100)
        (7,0.54083*100)
        (8,0.5243333*100)
        (9,0.5225*100)
        (10,0.519833*100)
        (11,0.52066666*100)
        (12,0.51933*100)
        (13,0.52833*100)
        (14,0.521666*100)
        (15,0.523333*100)
      };
         
    %German
    \addplot+[sharp plot,mark=star, clr2] coordinates
      {
      (0,0.91316*100)
        (1,0.7865*100)
        (2,0.73283*100)
        (3,0.6895*100)
        (4,0.6615*100)
        (5,0.62466*100)
        (6,0.6111666*100)
        (7,0.591333*100)
        (8,0.575666*100)
        (9,0.564333*100)
        (10,0.56916666*100)
        (11,0.56966666*100)
        (12,0.56033*100)
        (13,0.5516666*100)
        (14,0.5398333*100)
        (15,0.5355*100)
      };

    %Italian
    \addplot+[sharp plot,mark=|, clr3] coordinates
      {
      (0,0.97699*100)
        (1,0.6425*100)
        (2,0.608833*100)
        (3,0.571833*100)
        (4,0.5515*100)
        (5,0.5378*100)
        (6,0.525*100)
        (7,0.52033*100)
        (8,0.5085*100)
        (9,0.5138333*100)
        (10,0.51416*100)
        (11,0.50966*100)
        (12,0.517*100)
        (13,0.51433*100)
        (14,0.5106*100)
        (15,0.51266*100)
      };

     %Polish
    \addplot+[sharp plot, mark=square] coordinates
      {
        (0, 0.9436666*100)
        (1, 0.8183*100)
        (2, 0.746*100)
        (3, 0.6925*100)
        (4, 0.635666666*100)
        (5, 0.5918333*100)
        (6, 0.560833*100)
        (7, 0.553666*100)
        (8, 0.543*100)
        (9, 0.5335*100)
        (10, 0.5265*100)
        (11, 0.518*100)
        (12, 0.534*100)
        (13, 0.52366666*100)
        (14, 0.515*100)
        (15, 0.5063333*100)
      };

    %Spanish
    \addplot+[sharp plot, clr5] coordinates
      {
        (0, 0.985*100)
        (1, 0.7361*100)
        (2, 0.6926*100)
        (3, 0.6486666*100)
        (4, 0.612833*100)
        (5, 0.589*100)
        (6, 0.5705*100)
        (7, 0.54783*100)
        (8, 0.537333*100)
        (9, 0.5283333*100)
        (10, 0.524833*100)
        (11, 0.52583333*100)
        (12, 0.518*100)
        (13, 0.51366666*100)
        (14, 0.5055*100)
        (15, 0.5153333333*100)
      };

  \end{axis}
\end{tikzpicture}
% \end{document}
    \caption{Classifying inanimate grammatically feminine and masculine nouns approaches random guessing accuracy with each iteration of disentangling grammatical gender signals.}
    \label{fig:acc}
\end{figure}

\subsection{Grammatical Gender Signal Disentanglement}
Once grammatical gender signal ($\vec{d_g}$) is captured, it needs to be projected out of embedding $\vec{w}$ so that the resulting orthogonal embedding $\vec{w'}$ carries a reduced amount of grammatical gender signal as shown in Equation~\ref{eq:project-out}:

\begin{equation}
\vec{w'}= \vec{w}-\langle \vec{w}, \vec{d_g} \rangle \vec{d_g}
    \label{eq:project-out}
\end{equation}

Where $\langle \vec{x},\vec{y}\rangle$ is the inner product of $\vec{x}$ and $\vec{y}$. Note that although the same magnitude of grammatical gender is being projected out of all words, this process will impact embeddings differently; In an extreme example, if a word does not carry any grammatical gender signal component in its embedding (is orthogonal to $\vec{d_g}$), then $\vec{w'} = \vec{w}$.

While \citet{zhou-etal-2019-examining} apply LDA once to obtain the representation of grammatical gender, our method suggests that the hyperplane obtained using this technique ($\vec{d_g}$) does not fully capture grammatical gender signals: after $\vec{d_g}$ is projected out of embeddings, $\vec{w'}$ still carries significant grammatical gender signal as classifiers are still able to classify the grammatical gender of inanimate nouns with high accuracy. Thus, we iteratively extract the grammatical gender hyperplane and project it out of embeddings until the classifier reaches $50\%$ random-guessing accuracy in a binary task. 

\begin{figure}
    \centering
    \includegraphics[width=38mm]{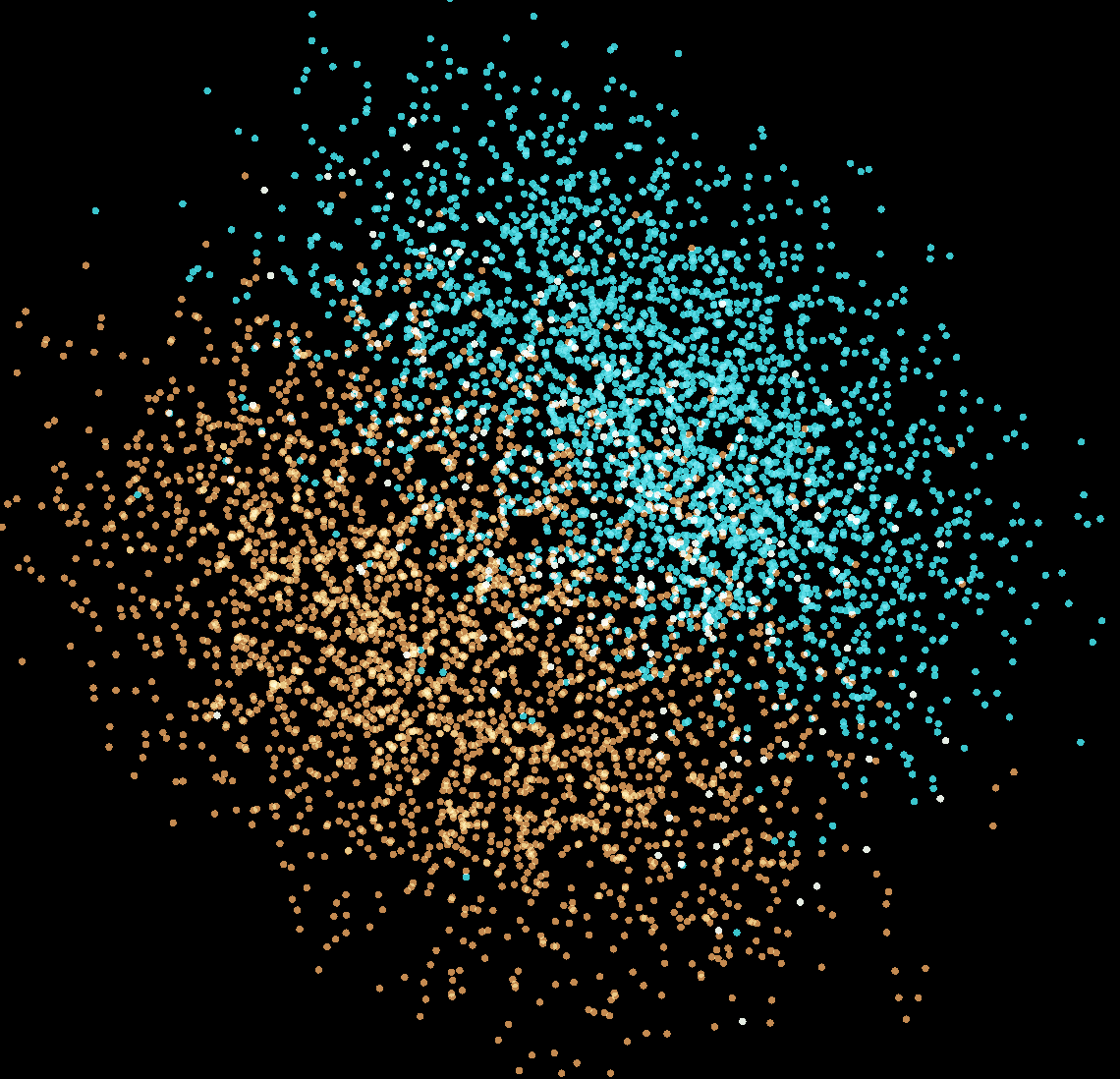}
    \includegraphics[width=40mm]{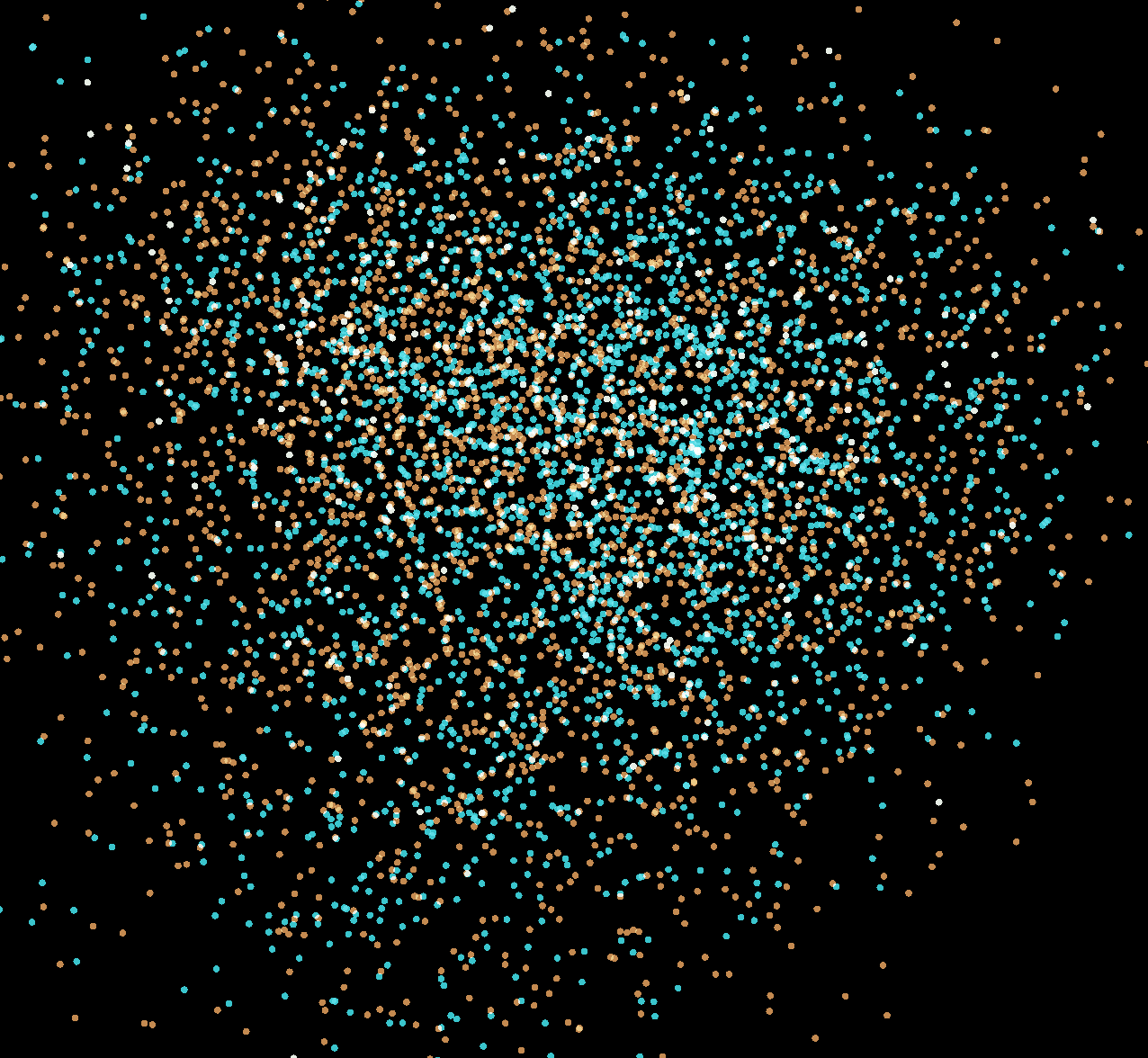}

    \caption{Visualization of Spanish word embeddings on two principal components before (left) and after (right) grammatical gender signal disentanglement.  Brown represents grammatically masculine, turquoise represents grammatically feminine nouns. Grammatical gender signals become random as visualized.}
    \label{fig:beforeafter}

\end{figure}

Initially, SVC models achieve an accuracy of above $96\%$ at classifying grammatically masculine and feminine embeddings for all languages, confirming our hypothesis that grammatical gender is indeed learned by word embeddings. Figure~\ref{fig:acc} shows classification accuracy during $15$ iterations of disentangling grammatical gender highlighting that after one iteration, classification accuracy is at an average of $75\%$. Figure~\ref{fig:beforeafter} demonstrates that before disentangling grammatical gender, Spanish inanimate feminine and masculine nouns are clearly separable. However, once disentanglement is complete, the nouns are much more intertwined in space and are no longer distinguishable.

\subsection{WEAT}
Biases in word embeddings before and after grammatical gender disentanglement are measured using WEAT. We present technical details for the original WEAT provided by \citet{Caliskan183} since it is the foundation of the different variations of WEAT that we use in our experiments:

WEAT quantifies the differential association between two target sets $X$ and $Y$ with two attribute sets $A$ and $B$. The output of WEAT is Cohen's effect size $d$ which measures the standardized magnitude of association:
\begin{equation}
d = \frac{mean_{x \in X} s(x, A, B) - mean_{y \in Y}s(y, A, B)}{std-dev_{w \in X \cup Y}s(w, A, B)}
\end{equation}
Where $s(w, A, B)$ measures the differential association of word $w$ with attribute sets $A$ and $B$ and is defined as:

\begin{equation}
s(w,A,B) = \scriptstyle{mean}_{a \in A} \scriptstyle{cos}(\vec{w}, \vec{a}) - \scriptstyle{mean}_{b \in B} \scriptstyle{cos}(\vec{w}, \vec{b})
\end{equation}

Note that $w$ refers to a word, $\vec{w}$ is the corresponding word embedding and $cos(\vec{a}, \vec{b})$ refers to the cosine similarity between the vectors $\vec{a}$ and $\vec{b}$. Additionally, the test statistic to measure the differential association of the target sets $X$ and $Y$ with the attribute sets $A$ and $B$ is computed the following way: 
\begin{equation}
   s(X,Y,A,B) = \sum_{x \in X} s(x,A,B) - \sum_{y \in Y} s(y,A,B) 
\end{equation}

While WEAT measures the differential association between two target sets and two attribute sets, single category WEAT (SC-WEAT)\footnote{\citet{Caliskan183} refers to single-category WEAT as WEFAT in their original work.} measures the differential association between a single word and two attribute sets. For example, SC-WEAT can measure whether the single word ``doll'' is more associated with a set of feminine words like ``mother'' and ``girl'' or a set of masculine words like ``father'' and ``boy''. Similar to WEAT, Cohen's effect size $d$ which measures the magnitude of association is computed as follows for SC-WEAT:
\begin{equation}
  d = \frac{\textrm{mean}_{a \in A} \textrm{cos}(\vec{w}, \vec{a}) - \textrm{mean}_{b \in B} \textrm{cos}(\vec{w}, \vec{b})}{\textrm{std-dev}_{x \in A \cup B}\textrm{cos}(\vec{w}, \vec{x})}   
\end{equation}

For both WEAT and SC-WEAT, Cohen's effect size($d$) quantifies the extent to which the association between the target(s) and attributes are separated from one another. Thus, a larger effect size signals a stronger magnitude of bias. $d$ values larger than $0.8$, $0.5$ and $0.2$ denote large, medium, and small effect sizes respectively. The experiments are designed so that a positive \textit{d} suggests a stereotype-congruent result while a negative value denotes a stereotype-incongruent result.

The significance of associations in WEAT and SC-WEAT is computed using a permutation test, which measures the (un)likelihood of the assumption that no associations exist between the target(s) and the attributes. With ${(X_i,Y_i)}_i$ denoting all partitions of $X \cup Y$ into two sets of equal size, the one-sided $p$ value of the permutation for WEAT is:
\begin{equation}
Pr_i[s[(X_i, Y_i, A, B) > s(X,Y,A,B)]
\end{equation}
The one-sided $p$ value of the permutation for SC-WEAT is:
 \begin{equation}
Pr_i[s(\vec{w, A_i, B_i}) > s(\vec{w,A,B})]
\end{equation}
Note that each word set in WEAT and SC-WEAT contains at least $8$ words to satisfy concept representation significance. Accordingly, the limitations of not adhering to this methodological robustness rule of WEAT, which are analyzed by \citet{ethayarajh2019understanding}, are mitigated.

\section{Experiments and Results}
\label{sec:experiments}

In section \ref{sec:approach}, it was demonstrated that grammatical gender signals are indeed learned by embeddings, and a method for disentangling grammatical gender was presented. In this section, we provide the details of a series of experiments that aim to answer our $3$ original questions: 
\begin{enumerate}
    \item Are social gender bias measurements impacted by grammatical gender? Is grammatical gender signal causing anomalous gender bias measurements?
    \item How does disentangling grammatical gender from inanimate nouns impact their semantic quality?
    \item Is grammatical gender effectively disentangled from word embeddings?
\end{enumerate}
 
The experiments detailed in subsections~\nameref{sec:soc_eval},~\nameref{sec:semqual}, and~\nameref{sec:gg-eval} correspond to questions 1,2 and 3.

\subsection{Social Gender Bias Evaluation}
\label{sec:soc_eval}
To understand the impact of grammatical gender on social gender bias measurements, we conduct two WEAT tests for measuring social (stereotypical) gender-bias which we collectively refer to as Gender WEAT:
\begin{enumerate}
    \item \textit{gender-science (GenS)} measures men:science-women:\\humanities association
    \item \textit{gender-career (GenC)} measures common names for boys:career-common names for girls:family association.
\end{enumerate}
GenS for Polish embeddings is the test for which \citet{toney2020valnorm} report anomalous gender bias measurements. If grammatical gender signals are causing anomalous gender WEAT measurements, we expect social gender bias scores after disentangling grammatical gender to match the ground truth gender bias scores. But what constitutes ground truth for gender WEAT experiments?

Since the WEAT experiments are adaptations of the IAT tests, comparison of the WEAT bias measurements to ground truth IAT scores as measured by social psychologists in various studies provides a proxy for validation of WEAT results. IAT measurements are used for establishing “stereotype-congruence” for each WEAT test; are men stereotypically more associated with career and women with family or vice versa according to IAT measurements? Obtaining a stereotype-incongruent WEAT result means that word embedding associations were different from IAT associations. For example, if IAT reports that human subjects tend to associate men with sciences and women with humanities but WEAT reports the opposite association, the WEAT result is said to be stereotype-incongruent.

GenS WEAT results are compared against the corresponding IAT results at the country level as reported by \citet{Nosek10593}, and GenC WEAT results are compared to IAT results at the language level as collected by \citet{osf} and computed by \citet{PMID:32747806}. We expect congruence (i.e match in sign), and not a one to one match in the magnitude of effect size. This is because WEAT and IAT effect sizes are not directly comparable, since IAT measures implicit biases of humans at individual level while word embeddings capture the aggregate biases in the statistical regularities of human-produced corpora \citep{Caliskan183}.

Table~\ref{tab:gender_weat} shows Gender WEAT results before and after disentangling grammatical gender for all languages in our study. Initially, Polish and Spanish GenS experiments display a men:humanities-women:science association, suggesting that grammatical gender signals may be stronger than potential social gender bias in these languages. However, the association becomes gender neutral ($d \approx 0$) for both languages afterwards (discussed in Section~\ref{sec:dis}). All other results are positive as they suggest stereotype-congruent associations. In all gender WEAT experiments, stereotypical gender biases reveal themselves more strongly as shown by the $\Delta$ column after disentangling grammatical gender. Thus, one can conclude that the presence of grammatical gender signals do impact social gender bias measurements. Yet, since we do not find evidence of stereotypical bias in Polish and Spanish GenS even after disentangling grammatical gender, it is not clear whether grammatical gender alone causes anomalous results. However, considering the five grammatical gender classes in Polish, it is possible that binary gender treatment of our method is not sufficiently disentangling grammatical gender signals in this language.

\begin{table}[]
    \centering
    \begin{tabular}{cccccccc}
         \hline
         \textbf{Lang} & \textbf{WEAT} & \pmb{$d_{init}$} & \pmb{$d_{SVC}$} & \pmb{$p_{SVC}$} & \pmb{$\Delta$} & \pmb{$d_{IAT}$} & \pmb{$C\textsuperscript{*}$}\\
        \midrule
        EN &  GenS & 0.88 & -- & -- &  -- & 0.38 & -- \\   
        & GenC & 1.66 & -- & -- & -- & 0.38 & -- \\
        \hline
        FR & GenS & 0.30 & 0.43 & 0.10 & +0.13 & 0.42 & \checkmark\\
        & GenC & 0.77 & 1.14 & 0.01 & +0.37 & 0.38 & \checkmark \\
        \hline
        DE &  GenS & 0.50 & 0.69 & 0.02 & +0.19 &0.42 & \checkmark\\
        & GenC  & 0.85 & 0.91 & 0.03 &  +0.06 & 0.40 & \checkmark\\
        \hline
        IT & GenS & -0.01 & 0.35 & 0.15 & +0.36 & 0.39 & \checkmark \\
        & GenC & 0.75 & 0.91 & 0.03 & +0.16 & 0.41 & \checkmark \\
        \hline
        PL & GenS & -0.26 & -0.07 & 0.58 & +0.19 & 0.49 & -- \\
        & GenC & 0.59 & 0.91 & 0.03 & +0.32 & 0.36 & \checkmark \\
        \hline
        ES & GenS & -0.32 & 0.04 & 0.46 & +0.36 & 0.33 & -- \\ 
        & GenC & 1.31 & 1.52  & $10^{-3}$ &  +0.21 & 0.35 & \checkmark \\
        \hline 
    \end{tabular}
    \caption{Gender WEAT results before and after disentangling grammatical gender with SVC. $d_{init}$ denotes the initial effect size (magnitude of bias measured in Cohen's $d$), and $d_{SVC}$ refers to effect size after disentangling grammatical gender. $p$ is a measure of statistical significance, $\Delta$ measures the change in effect size ($d_{SVC}-d_{init}$). $d_{IAT}$ denotes the ground truth effect sizes reported by \citep{PMID:32747806} for GenC and \citep{Nosek10593} for GenS. \pmb{$ C\textsuperscript{*}$} stands for congruent, which denotes whether $d_{SVC}$ matches in sign with $d_{IAT}$.}
    \label{tab:gender_weat}
\end{table}

\subsection{Semantic Quality Evaluation}
\label{sec:semqual}
In this section we conduct three different experiments to evaluate the impact of grammatical gender disentanglement on the semantic quality of the embeddings, including semantic gender information:

\subsubsection{Analogy Tests} 
\label{sec:analogy}
If we have disentangled only grammatical gender information, and no semantic gender information along with it, we expect no change in embeddings' ability to solve analogies such as \textit{man:woman-nephew:niece}. Such word analogies are common embedding intrinsic evaluation tasks \cite{mikolov2013efficient}.\footnote{We use analogy datasets for French and Polish \cite{grave2018learning}, German \cite{koper-etal-2015-multilingual}, Italian \cite{berardi2015word}, and Spanish \cite{cardellinoSBWCE}} We choose the {\fontfamily{qcr}\selectfont family} subset of the analogy dataset as it specifically tests for semantic gender quality of embeddings. We additionally select the {\fontfamily{qcr}\selectfont capital-common-countries} subset of the analogy dataset because it is another task that evaluates semantic quality as opposed to syntactic. Table~\ref{tab:analogy} shows performance on the combined analogy tasks. There is a nontrivial improvement in analogy performance for German, Italian and Polish as words are no longer affected by the grammatical gender signal. \citet{gonen-etal-2019-grammatical-gender} also suggest that disentangling grammatical gender can improve the semantic quality of embeddings.
\begin{table}[h]
\begin{tabular}{cccc}
\hline {\textbf{Lang}} & {\pmb{$\alpha_i$ (\%)}} & {\pmb{$\alpha_a$ (\%)}} & {\pmb{$N$}}\\ \midrule
{EN} & {80} & {--} & {1,012}\\
{FR} & {62} & {62} & {722}\\
{DE} & {67} & {68} & {1,012} \\
{IT} & {72} & {73} & {848}\\
{PL} & {54} & {57} & {926}\\
{ES} & {82} & {82} & {800} \\
\hline
\end{tabular}
    \caption{\label{tab:analogy} Accuracy (shown in percentage \%) in the analogy test before (\textbf{$\alpha_i$}) and after (\textbf{$\alpha_a$}) disentangling grammatical gender. \textbf{$N$} denotes the number of analogies.}
\end{table}

\subsubsection{ValNorm}
\label{sec:valnorm}
Another method of ensuring semantic utility preservation after disentangling grammatical gender is ValNorm \citep{toney2020valnorm} which is a word embedding intrinsic evaluation task that measures the alignment of human valence (pleasantness/unpleasantness) norm scores with valence associations in embeddings. ValNorm is designed for evaluating the semantic representativeness and quality of embeddings when measuring bias, as it captures widely accepted consistent non-social group associations in various languages. In this benchmark, Pearson's correlation coefficient between the human judgement scores of valence for $399$ words and the valence associations in word embeddings is validated. Valence datasets are detailed in the appendix. Table~\ref{tab:valnorm} summarizes the results which confirm that semantic associations are improved in French, Polish, and Spanish and preserved for other languages after disentangling grammatical gender.
\begin{table}[h]
\begin{tabular}{cccc}
\hline {\textbf{Lang}} & {\pmb{$\rho_i$}} & {\pmb{$\rho_a$}} & {\pmb{$N$}}\\ \midrule
{EN} & {0.87} & {--} & {381} \\
{FR} & {0.82} & {0.83} & {354}\\
{DE} & {0.78} & {0.78} & {370} \\
{IT} & {0.82} & {0.82} & {374}\\
{PL} & {0.66} & {0.71} & {365}\\
{ES} & {0.81} & {0.82} & {381} \\
\hline
\end{tabular}
\caption{\label{tab:valnorm} Pearson's correlation coefficient before (\textbf{$\rho_i$}) and after (\textbf{$\rho_a$}) disentangling grammatical gender signals. \textbf{$N$} denotes the number of words in ValNorm. Higher \textbf{$\rho$} suggests embeddings more aligned with human valence norm scores.}
\end{table}

\subsubsection{Baseline WEAT} 
Baseline WEAT are among the original WEAT variations used by \citet{Caliskan183} to quantify widely shared baseline associations such as relatively positive attitude towards flowers and negative attitude towards insects. These associations are expected to manifest in different languages or populations \cite{pmid9654756}. Consequently, we use the following widely shared association tests as a baseline to evaluate the semantic representation accuracy of word embeddings in any language:
\begin{enumerate}
    \item \textit{flowers-insects (FloI)} measures pleasant:flowers-unpleasant:\\insects association
    \item \textit{instruments-weapons (InsW)} measures pleasant:musical \\instruments-unpleasant:weapons association
\end{enumerate}
We do not expect grammatical gender to have a significant impact on baseline WEAT as it does not measure gender related associations. Thus, they are included as a baseline where we expect high effect size both before and after disentangling grammatical gender. Despite expecting little change in baseline associations, there is a remarkable improvement in Polish baseline associations suggestive of a potential semantic quality improvement as shown in Table~\ref{tab:main-results} (All results in the paper are statistically significant unless noted otherwise). This is in line with observations from analogy and ValNorm experiments, where Polish embeddings had the largest magnitude of semantic improvement.

The results of analogy task, ValNorm and baseline WEAT all suggest that not only the semantic quality of embeddings is preserved after disentangling grammatical gender, but also the semantic quality of word embeddings may have slightly improved, especially for Polish embeddings. 
\begin{table}[h!]
\begin{center}
\centering
\begin{tabular}{c|c|ccccc}
\hline
\textbf{Test}& \textbf{Lang} & \textbf{WEAT} & \pmb{$d_{init}$} & \pmb{$d_{SVC}$} & \pmb{$p_{SVC}$} & \pmb{$\Delta$} \\
\midrule
Baseline & EN & FloI & 1.45 & -- & -- & -- \\ \cline{3-7}
& &InsW & 1.54 & -- & -- & -- \\ 
\cline{2-7}
& FR & FloI & 1.38 & 1.44 & $10^{-3}$ & +0.06 \\ \cline{3-7}
&& InsW & 1.40 & 1.45 & $10^{-6}$ & +0.05 \\ 
\cline{2-7}
&DE &  FloI & 1.58 & 1.62 & $10^{-5}$ & +0.04 \\ \cline{3-7}
& & InsW & 1.63 & 1.67 & $10^{-9}$ & +0.04 \\ 
\cline{2-7}

& IT & FloI & 1.56 & 1.61 & $10^{-4}$ & +0.05 \\ \cline{3-7}
& & InsW & 1.42 & 1.37 & $10^{-6}$ & -0.05\\ 
\cline{2-7}

& PL & FloI &  1.00 & 1.34 & $10^{-3}$ & +0.34\\ \cline{3-7}
& & InsW &  0.71 & 1.19 & $10^{-5}$ & +0.48 \\ 
\cline{2-7}

& ES & FloI & 1.67 & 1.60 & $10^{-4}$ & -0.07 \\ \cline{3-7}
& & InsW & 1.08 & 1.12 & $10^{-4}$ & +0.04 \\ 
\hline 
Grammatical & FR & GG & 1.82 & 0.40 & 0.20 &  -1.42 \\ 
\cline{2-7}

Gender &DE &  GG & 1.55 & 0.30 & 0.26 & -1.25 \\ 
\cline{2-7}

Evaluation & IT & GG & 1.79 & 0.58 & 0.11 & -1.21\\ 
\cline{2-7}

& PL & GG &  1.85 & 1.34 & $10^{-3}$ & -0.51\\ 
\cline{2-7}

& ES & GG & 1.83 & 1.23 & $10^{-3}$ & -0.60 \\ 
\hline 
\end{tabular}
\caption{\label{tab:main-results}
{Baseline and evaluation WEAT results before and after disentangling grammatical gender with SVC. $d_{init}$ denotes the initial effect size (magnitude of bias measured in Cohen's $d$), and $d_{SVC}$ refer to effect size after disentangling grammatical gender. $p$ is a measure of statistical significance, $\Delta$ measures the change in effect size ($d_{SVC}-d_{init}$).}} 
\end{center}
\end{table}
\subsection{Grammatical Gender Disentanglement Evaluation}
\label{sec:gg-eval}
In this section, we conduct three different experiments to evaluate the effectiveness of our proposed grammatical gender disentanglement method:

\subsubsection{GG-WEAT}

GG-WEAT evaluates whether the association between inanimate nouns and semantically gendered words weakens after disentangling grammatical gender. The attribute sets in GG-WEAT are words with feminine and masculine semantic gender, such as \{mother, daughter\} vs. \{father, son\}. The target sets are inanimate grammatically feminine vs. masculine nouns. In the absence of grammatical gender, random inanimate nouns should be equally similar to words like "mother" and "father" regardless of their grammatical gender (disregarding social gender bias). In the presence of grammatical gender, we expect association between inanimate feminine/masculine nouns with feminine/masculine words. Thus, by measuring GG-WEAT effect size, one can evaluate the presence and strength of grammatical gender in the embedding space since higher effect sizes indicate stronger grammatical gender presence.

\begin{figure}
\centering
    % \documentclass[border=10pt]{standalone}
% \usepackage{pgfplots}
% \usepgfplotslibrary{fillbetween}
% \begin{document}
\begin{tikzpicture}
\pgfplotsset{height=6cm, width=63mm}
\definecolor{clr1}{cmyk}{100,0,0,0}
\definecolor{clr2}{cmyk}{0,100,0,0}
\definecolor{clr3}{cmyk}{0,50,100,0}
\definecolor{clr4}{cmyk}{100,69,0,0}
\definecolor{clr5}{cmyk}{60,14,68,0}

  \begin{axis}[
  title= GG-WEAT Effect Size vs. Iteration,
  xlabel={Number of Iterations},
  ylabel={GG-WEAT Effect Size (Cohen's $d$)},
  grid=major,
  legend entries={FR,DE,IT,PL,ES}, legend style={at={(1.35,1.0)}, font=\tiny},
  ymin=0,
  ymax=2,xmin=0,
  xmax=15,]
  
    \addplot+[mark=triangle, sharp plot, clr1] coordinates
    %french
      {
       (0,1.82)
        (1,1.38)
        (2,1.21)
        (3,1.09)
        (4,0.98)
        (5,0.94)
        (6,0.9)
        (7,0.86)
        (8,0.75)
        (9,0.68)
        (10,0.66)
        (11,0.65)
        (12,0.62)
        (13,0.58)
        (14,0.57)
        (15,0.57)
        % (16,0.55)
        % (17,0.56)
        % (18,0.55)
        % (19,0.54)
        % (20,0.52)
      };
         
    %German
    \addplot+[sharp plot,mark=star, clr2] coordinates
      {
        (0,1.55)
        (1,1.04)
        (2,0.74)
        (3,0.58)
        (4,0.5)
        (5,0.4)
        (6,0.32)
        (7,0.27)
        (8,0.32)
        (9,0.32)
        (10,0.32)
        (11,0.32)
        (12,0.32)
        (13,0.32)
        (14,0.31)
        (15,0.3)
        % (16,0.3)
        % (17,0.3)
        % (18,0.28)
        % (19,0.3)
        % (20,0.29)
      };

    %Italian
    \addplot+[sharp plot,mark=|, clr3] coordinates
      {
        (0,1.79)
        (1,1.52)
        (2,1.44)
        (3,1.38)
        (4,1.38)
        (5,1.38)
        (6,1.38)
        (7,1.37)
        (8,1.36)
        (9,1.35)
        (10,1.33)
        (11,1.33)
        (12,1.28)
        (13,1.22)
        (14,1.11)
        (15,1.01)
        % (16,0.95)
        % (17,0.87)
        % (18,0.81)
        % (19,0.81)
        % (20,0.8)
      };

     %Polish
    \addplot+[sharp plot, mark=square] coordinates
      {
        (0,1.85)
        (1,1.77)
        (2,1.76)
        (3,1.75)
        (4,1.73)
        (5,1.73)
        (6,1.71)
        (7,1.67)
        (8,1.59)
        (9,1.56)
        (10,1.51)
        (11,1.46)
        (12,1.44)
        (13,1.41)
        (14,1.39)
        (15,1.36)
        % (16,1.34)
        % (17,1.34)
        % (18,1.33)
        % (19,1.34)
        % (20,1.34)
      };

    %Spanish
    \addplot+[sharp plot, clr5] coordinates
      {
        (0,1.83)
        (1,1.55)
        (2,1.51)
        (3,1.47)
        (4,1.45)
        (5,1.43)
        (6,1.42)
        (7,1.42)
        (8,1.42)
        (9,1.42)
        (10,1.42)
        (11,1.42)
        (12,1.41)
        (13,1.38)
        (14,1.34)
        (15,1.3)
        % (16,1.29)
        % (17,1.27)
        % (18,1.26)
        % (19,1.25)
        % (20,1.25)
      };

  \end{axis}
\end{tikzpicture}
% \end{document}
    \caption{Change in GG-WEAT effect size (measured in Cohen's $d$) with iterations of grammatical gender disentanglement.}
    \label{fig:gg}
\end{figure}
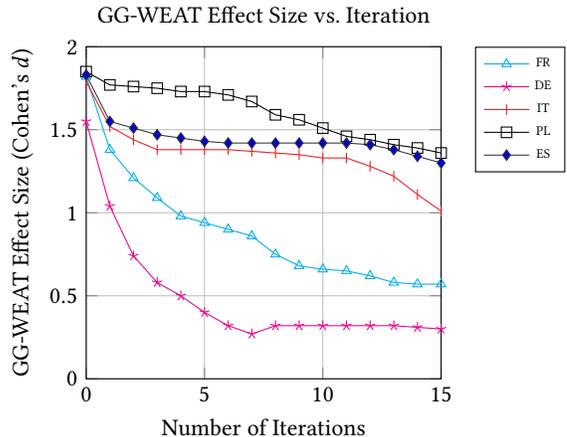

If grammatical gender is indeed being disentangled, we expect to see a decrease in effect size \textit{d} and loss of statistical significance with each iteration of disentangling grammatical gender. Figure~\ref{fig:gg} shows that initially, the grammatical gender effect sizes are very high (larger than $1.5$ for all languages). Even after one iteration of disentanglement, the signal is a strong $d>1.0$ for all languages, emphasizing the need for the iterative process. After disentangling grammatical gender iteratively the effect sizes for FR, DE, and IT become small ( $p_{SVC}$ is no longer statistically significant as shown in Table~\ref{tab:main-results}) suggesting that grammatical gender signals have been successfully disentangled to a great degree. However, despite a decrease in magnitude, final effect sizes for PL and ES are still high and $p_{SVC}$ is still significant. These are also the same languages for which we did not find stereotype-congruent GenS results. This could suggest that we are only partially disentangling grammatical gender for Spanish and Polish, and a complete disentanglement requires more complicated methodologies tailored for language structure.

In order for GG-WEAT to quantify grammatical gender signals, the target sets in GG-WEAT must be constructed in a way such that they have minimal difference in their stereotypical association with a semantic gender. Otherwise semantic gender bias would interfere with grammatical gender measurements. For example, although \textit{la moda} and \textit{el baloncesto} are inanimate feminine and masculine nouns, we do not want to include them in GG-WEAT target sets because "fashion" and "basketball" have stereotypical feminine and masculine associations, respectively. Thus, even after disentangling grammatical gender, according to current stereotypes we expect association between \textit{moda} and femininity and \textit{baloncesto} with masculinity to persist.  

To control for this, we construct the target set by taking pairs of inanimate nouns with a high semantic similarity score  which have opposite grammatical gender from the Simlex-999 dataset \citep{10.1162/COLI_a_00237}.\footnote{We use Simlex-999 translations of \citet{barzegar-etal-2018-semr} for Spanish and French, \citet{leviant2015separated} for German and Italian and \citet{MYKOWIECKA18.687} for Polish.} For example, feminine \textit{molecula} and masculine \textit{atomo} are placed in opposite target sets for having a high similarity score in Italian Simlex-999 (and thus for not being much different in their relative association to semantically feminine and masculine concepts).

\subsubsection{Single-category GG-WEAT}
A single-category variation of GG-WEAT (consisting of one inanimate word as the target and semantically feminine and masculine words as attribute sets) allows for the quantification of grammatical gender association at the word level. For example, the effect size for \textit{fuerza} the grammatically feminine Spanish word for strength is initially $1.0$ suggesting a high association with semantic femininity\footnote{Let $d$>0 denote association with semantic femininity, and $d$ <0 with semantic masculinity.} (despite strength being a stereotypically masculine trait). We re-compute effect size after disentangling grammatical gender to observe whether any stereotypical masculine association is hidden under the impact of grammatical gender. 

\begin{figure}
    \centering
 \includegraphics[scale=0.6]{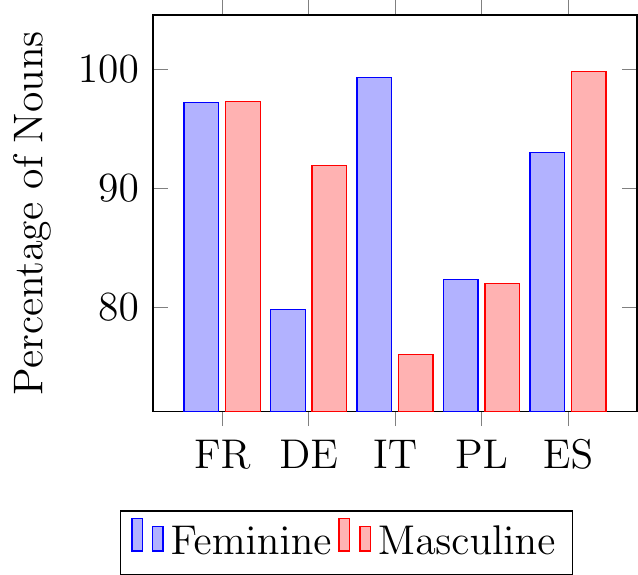}
    \caption{The bar chart shows the percentage of inanimate nouns whose association with their grammatical gender weakened after disentangling grammatical gender.}
    \label{fig:barchart}
\end{figure}

\begin{figure}
    \centering
    \includegraphics[scale=0.20]{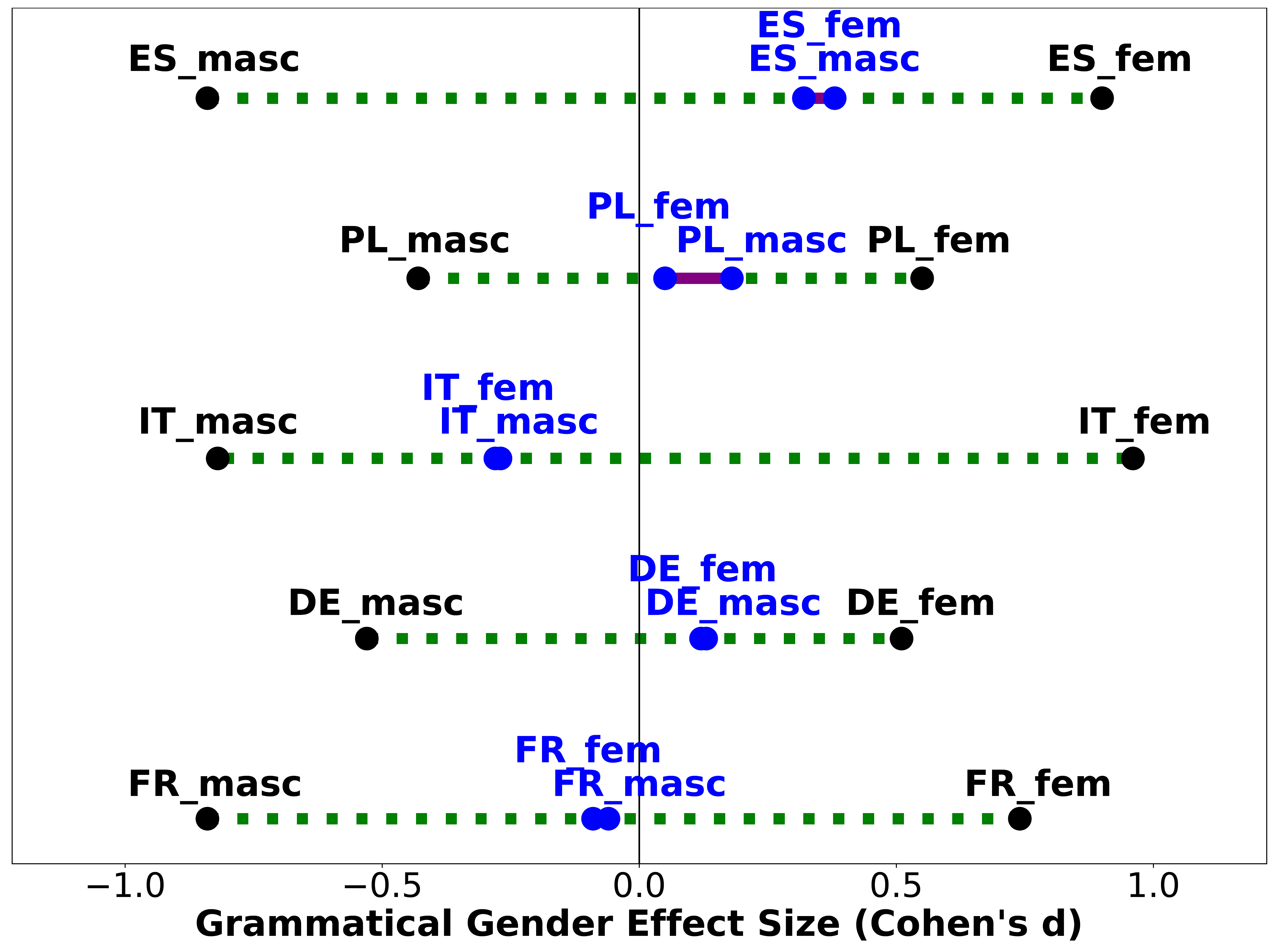}
    \caption{Black and blue dots represent the average GG-WEAT magnitude before and after disentangling grammatical gender for $2,000$ inanimate masculine and feminine nouns. The dotted green lines between black and the solid purple lines between the blue points show the magnitude of difference before and after disentangling grammatical gender. On average, inanimate nouns become more grammatically gender neutral and the difference between masculine and feminine nouns becomes smaller.}
    \label{fig:sc-gg}
\end{figure}

We perform single-category GG-WEAT for $2,000$ inanimate grammatically feminine and masculine nouns per each language and find that the effect size for over $90\%$ of the words moves toward the gender-neutral direction as shown in Figure~\ref{fig:barchart}. The semantic gender association gap between grammatically feminine and masculine nouns becomes significantly smaller after disentangling grammatical gender and both sets move much closer to gender neutrality as shown in Figure~\ref{fig:sc-gg}. The effect size for our motivating word \textit{fuerza} (Spanish for strength), which used to be highly associated with femininity because of its grammatical gender, becomes gender neutral ($d = -0.01$; showing no sign of stereotypical masculine association after disentangling grammatical gender).

\subsubsection{Pairwise Distance}
To evaluate the neutralization of grammatical gender, \citet{gonen-etal-2019-grammatical-gender} split the inanimate portion of Italian and German Simlex-999 datasets in two sets. First set includes pairs of words with the same grammatical gender, and second set includes pairs of words with opposite gender.\footnote{In this method, there could be imbalance between the sizes of the two sets. For example, there are $404$ pairs of words with different genders in German dataset while only $124$ pairs of words have the same gender.} The corresponding English sets for each language are also created. The average cosine similarity between pairs of words with same gender $avg_{s}$ and different gender $avg_{d}$ are compared. If grammatical gender impacts word embeddings, then on average one expects pairs of words with the same gender to be more similar to one another compared to pairs of words with the opposite gender (i.e $avg_{s} > avg_{d}$). Corresponding English embeddings approximate what the difference between the two similarities should be without the impact of grammatical gender. Let $l_{e} = avg_{s} - avg_{d}$ for English, and $l'_{g} = avg_{s} - avg_{d}$ denote the difference for gendered language after disentangling grammatical gender. Let $l_{g}$ denote the original difference for the gendered language. Grammatical gender signal reduction is computed as shown in Equation~\ref{eq:reduction} and results are shown in Table~\ref{tab:gonen}.
 \begin{equation}
    reduction = 1- \frac{l'_{g} - l_{e}}{l_{g}- l_{e}}
    \label{eq:reduction}
\end{equation}

\begin{table}[ht]
\begin{center}
\begin{tabular}{ccccc}
\hline \small{FR} & \small{DE} & \small{IT} & \small{PL} & \small{ES}\\ \hline
\small{53\%} & \small{21\%} & \small{61\%} & \small{36\%} & \small{40\%}\\
\hline \\
\end{tabular}
\caption{\label{tab:gonen} Gap reduction (\%) denotes how much the difference between the average similarity of words with the same grammatical gender and  average similarity of words with different grammatical gender of a gendered language become more similar to its English translation.}  
\end{center}
\end{table}
While results in Table~\ref{tab:gonen} may suggest that German grammatical gender signal is the most resistant to disentanglement (due to smaller reduction), GG-WEAT suggests that Polish signal is the most resistant. However, note that although English embeddings (which are the basis of pairwise distance method) are a good estimation of how far apart the same gender and opposite gender pairs of words should be without the effect of grammatical gender, one cannot expect the semantic spaces of English and other languages to be exactly comparable \cite{thompson2020cultural}. GG-WEAT on the other hand does not rely on the embedding space of other languages to evaluate grammatical gender presence.

\section{Discussion}
\label{sec:dis}
Although IAT measurements report GenS, which measures the association between men with sciences and women with humanities \cite{Nosek10593} as a metric to measure social gender biases in human cognition, historically men have dominated the humanities in addition to science and engineering fields. Most notable philosophers, historians, linguists, writers, poets and artists throughout the history have been overwhelmingly men while women have always been more associated with domestic roles. Moreover, \citet{caliskan2022gender} provide evidence for a masculine default in English word embeddings, where various semantic domains are strongly associated with men. Therefore, it should not come as a surprise that GenS effect sizes are consistently smaller than GenC measurements. 

Furthermore, if the distribution of grammatical gender in a language is not balanced, the magnitude of grammatically feminine and masculine signals might be different. Accordingly, using the masculine and feminine version of a word might not normalize the grammatical gender signals. In our case, the grammatical gender subspace resulting from SVC may not contain feminine and masculine grammatical gender information at an equal rate. This may explain why after disentanglement, inanimate nouns still deviate from grammatical gender neutrality as shown in Figure~\ref{fig:sc-gg}. Overcoming this imbalance requires evaluating grammatical gender signals while taking other parameters such as word frequency and grammatical gender distribution in a language into account.

Our final GenS measurements in Spanish and Polish do not report stereotypical associations between men with sciences and women with humanities as reported by social psychologists\citep{Nosek10593}. However, according to Eurostat\footnote{\url{https://ec.europa.eu/eurostat/en/web/main/data/database}} gender gap in science has been decreasing over the past decade in the European Union. The percentages of female scientists and engineers were reported as $47\%$ and $46\%$ for Spain and Poland in year $2021$. Since IAT might be reflecting societal structure \citep{PayneB.Keith2017TBoC} and embedding associations correlate with real world gender statistics \citep{Caliskan183}, this approximately equal representation may explain our gender neutral results to some extent\footnote{Female scientists and engineers proportions for France, Italy, and Germany are $39\%$, $21\%$, and $37\%$ respectively which are smaller compared to Spain and Poland.  The dataset can be found \href{https://appsso.eurostat.ec.europa.eu/nui/show.do?query=BOOKMARK_DS-063435_QID_-23E982DF_UID_-3F171EB0&layout=TIME,C,X,0;GEO,L,Y,0;UNIT,L,Z,0;CATEGORY,L,Z,1;AGE,L,Z,2;SEX,L,Z,3;INDICATORS,C,Z,4;&zSelection=DS-063435UNIT,THS;DS-063435CATEGORY,SE;DS-063435SEX,T;DS-063435AGE,Y15-24_Y65-74;DS-063435INDICATORS,OBS_FLAG;&rankName1=UNIT_1_2_-1_2&rankName2=INDICATORS_1_2_-1_2&rankName3=CATEGORY_1_2_-1_2&rankName4=SEX_1_2_-1_2&rankName5=AGE_1_2_-1_2&rankName6=TIME_1_0_0_0&rankName7=GEO_1_2_0_1&sortC=ASC_-1_FIRST&rStp=&cStp=&rDCh=&cDCh=&rDM=true&cDM=true&footnes=false&empty=false&wai=false&time_mode=NONE&time_most_recent=false&lang=EN&cfo=\%23\%23\%23\%2C\%23\%23\%23.\%23\%23\%23}{at this link.}}.

Note that our method assumes the presence of only two grammatical gender classes in a language. However, Polish and German have an additional neuter class. It is not clear how associations from ``neuter'' class might be impacting bias measurements. Exploring more effective grammatical gender disentanglement methods as they relate to other representations of grammatical gender is left for future work. Similarly, extending the methods to non-binary or other representations of semantic gender is left to future work due to low signal strength for underrepresented groups.

Additionally, although results in Table~\ref{tab:gender_weat} suggest that disentangling grammatical gender from embeddings leads to obtaining social gender bias signals that are more congruent with country level bias statistics (i.e larger effect sizes in gender WEAT) this does not always have to be the case; the change in effect size could depend on factors such as the feminine/masculine ratio of the stimuli. In our experiments, the number of grammatically masculine words in humanities were consistently larger than grammatically masculine science  words which contributed to stereotype-incongruent associations. Thus, an increase in GenS effect size in stereotype-congruent direction is plausible. Yet, feminine/masculine ratios alone may not determine the direction of effect size's shift because the magnitude of grammatical gender learned by one word may be different from another due to factors such as frequency in training corpus.

Repeating GG-WEAT experiments with Word2Vec embeddings of \citet{fares-etal-2017-word} also suggest that Spanish and Polish grammatical gender signals are more resistant to our grammatical gender disentanglement method. However, fewer iterations are needed to achieve random guessing accuracy which is accompanied by a slight decrease in semantic quality of embeddings. This could be due to the difference in training algorithms as FastText embeddings are sub-word aware and grammatical gender marks the ending of many nouns. Yet, the different training conditions of the two sets of embeddings means that they are not directly comparable: The embeddings have been trained on different corpora of varying set size. Word2Vec embeddings of \citet{fares-etal-2017-word} have 100 dimensions whereas FastText embeddings are 300 dimensional. Embedding qualities are also very different. These reasons, along with other factors, make direct comparison of the embedding algorithm futile. Future research can focus on how grammatical gender manifests itself through different embedding training algorithms, including dynamic word embeddings generated by language models.

Finally, this work does not study if grammatical gender influences social bias in human cognition. Such an evaluation requires more fine-grained grammatical gender signal extraction approaches as well as different experimental settings.

\section{Conclusion}
Grammatical gender signals can interfere and potentially cause anomalous results when measuring social gender bias in word embeddings of gendered languages. The above $96\%$ accuracy in classifying inanimate feminine and masculine nouns in five gendered languages indicates that grammatical gender is indeed learned by word embeddings.
However, once grammatical gender is detected and disentangled from embeddings, the association between inanimate nouns and semantically gendered words measured by Cohen's $d$ reduces by an average of $d = 1.0$. After disentangling syntactic gender signals, stereotypical gender biases reveal themselves more strongly by an average of $d = 0.24$, and semantic quality in Polish embeddings is evidently improved. 

\section{Impact Statement}
Since practitioners use cross-lingual pre-trained word embeddings in different applications, bias in word embeddings may potentially propagate to downstream tasks. To uphold the notion of fairness, practitioners must be able to accurately measure such potential biases in embeddings and be aware that language structure can impact social bias measurements. Therefore, a hasty generalization of bias measurement tools from one language to another without taking language properties such as grammatical gender into account can lead to failure in accurate identification and prevention of bias. Disentangling grammatical gender is one method that can help with more accurate measurements of social biases in word embeddings.

\newpage
%%
%% The next two lines define the bibliography style to be used, and
%% the bibliography file.
\bibliographystyle{ACM-Reference-Format}
\bibliography{sample-base}

%%
%% If your work has an appendix, this is the place to put it.
\appendix
\section{WEAT Stimuli List}
Although science-humanities stimuli were present in Project Implicit Website for all languages of interest, we expanded the size of the stimuli by adding more subjects. These additional subjects were taken from ACTs list of college majors and occupational choices \footnote{\url{http://www.act.org/content/act/en/research/reports/act-publications/college-choice-report-class-of-2013/college-majors-and-occupational-choices/college-majors-and-occupational-choices.html}}. The intuition was that a larger number of words can represent a concept more accurately in word embeddings. For example, the word "English" alone cannot represent the concept of humanities/social science subjects, because it has many other senses beyond an academic subject such as referring to a person from England. The words "English" and "arts" together are better representatives of these subjects, but are not representative enough so even more words are needed. In short, stimuli expansion helps with more robust representation of concepts by word embeddings. Stimuli expansion was only needed for science-humanities because these concepts are much more overlapping and less mutually exclusive compared to other concepts. At the end, mathematics, natural sciences, humanities and social sciences are all academic disciplines, whereas concepts such as career-family are much more distinct.

Furthermore, whenever translating an English term resulted in a word with more than one part, we omit those words from the stimuli as our embeddings are tokenized and are not available for two part words. For example, translating the term "teargas" from English weapons stimuli to French results in \textit{gaz lacrymogène}, which does not have a word embedding. Additionally, if translating multiple English terms results in duplicate translations we only keep one of the words. For example, Google translate suggest the word \textit{matrimonio} as the translation of both "wedding" and "marriage" in GenC stimuli. We remove duplicate terms from the translation stimuli. 

The stimuli used for Gender WEAT is provided below. For the complete stimuli list (including baseline and GG-WEAT) please refer to the code.
Masculine nouns have been written in blue, and neuter nouns (present in German and Polish stimuli) are in purple. Nouns written in black are feminine (except for English, where there is no grammatical gender). Words that could be masculine or feminine or not gendered (such as adjectives or verbs in some languages) are marked with *.
\subsection{English}
\textit{GenS}\\
\textbf{science}: astronomy, math, chemistry, physics, biology, geology, engineering, statistics, bioengineering, biophysics, biochemistry, ecology, microbiology, algebra, geometry, telecommunications, computer, astrophysics\\
\textbf{humanities}: history, arts, humanities, english, philosophy, music, literature, psychology, sociology, geography, anthropology, theology, linguistics, journalism, archaeology, dancing, drawing, painting \\
\textbf{men}:man, son, father, boy, uncle, grandpa, husband, male \\
\textbf{women}: mother, wife, aunt, woman, girl, female, grandma, daughter \\
\textit{GenC}\\
\textbf{career}: career, corporation, salary, office, professional, management, business \\
\textbf{family}: wedding, marriage, parents, relatives, family, home, children \\
\textbf{men}: Ben, Paul, Daniel, John, Jeffrey \\
\textbf{women}: Rebecca, Michelle, Emily, Julia, Anna 

\subsection{French}
\textit{GenS}

\noindent\textbf{science}: astronomie, mathématiques, chimie, physique \footnote{If the word refers to physics as academic discipline, it is masculine. If it refers to body and physical appearance, it is feminine.}, biologie, géologie, ingénierie, statistiques, bioingénierie, biophysique, biochimie, écologie, microbiologie, algèbre, géométrie, télécommunications, \textcolor{blue}{ordinateur}, astrophysique\\
\textbf{humanities}: philosophie, humanités, \textcolor{blue}{art}, \textcolor{blue}{latin}, littérature, musique, histoire, psychologie, sociologie, géographie, anthropologie, théologie, linguistique, \textcolor{blue}{journalisme}, archéologie, danse, \textcolor{blue}{dessin}, peinture \\
\textbf{men}: \textcolor{blue}{garçon}, \textcolor{blue}{père}, \textcolor{blue}{masculin}, \textcolor{blue}{mari}, \textcolor{blue}{fils}, \textcolor{blue}{oncle}\\
\textbf{women}:  demoiselle, féminin, tante, fille, femme, mère\\
\textit{GenC}\\
\textbf{career}: carrière, corporation, \textcolor{blue}{salaire}, \textcolor{blue}{bureau}, professionnel*, gestion, entreprise\\
\textbf{family}: \textcolor{blue}{mariage}, \textcolor{blue}{domicile}, parents*, \textcolor{blue}{proches}, famille, maison, enfants* \\
\textbf{men}: \textcolor{blue}{Nicolas}, \textcolor{blue}{Alexandre}, \textcolor{blue}{Guillaume}, \textcolor{blue}{Mathieu}, \textcolor{blue}{Thomas}, \textcolor{blue}{Pierre}, \textcolor{blue}{Emmanuel}, \textcolor{blue}{Jean}, \textcolor{blue}{François} \\
\textbf{women}: Céline, Marie, Sandrine, Sophie, Caroline, Julie, Hélène, Camille, Emilie\\

\subsection{German}
\textit{GenS}

\noindent\textbf{science}: Astronomie, Mathematik, Chemie, Physik, Biologie, Geologie, Ingenieurswissenschaften, Statistik, Bioingenieurwesen, Biophysik, Biochemie, Ökologie, Mikrobiologie, Algebra, Geometrie, Telekommunikation, \textcolor{blue}{Computer}, Astrophysik

\noindent\textbf{humanities}: Philosophie, Kunst, Geschichte,Musik, Geisteswissenschaften, Psychologie, Soziologie, Geographie, Anthropologie, Theologie, Linguistik, \textcolor{blue}{Journalismus}, Archäologie, \textcolor{blue}{Tanz}, Zeichnung, Malerei, Sprachwissenschaften, Literaturwissenschaften\\
\textbf{men}: \textcolor{blue}{Mann, Junge, Vater, Männlich, Großvater, Ehemann, Sohn, Onkel}\\
\textbf{women}: Mädchen, Weiblich, Tante, Tochter, Ehefrau, Frau, Mutter, Großmutter\\
\textit{GenC}\\
\textbf{career}: Verwaltung, Berufstätigkeit, \textcolor{purple}{Unternehmen}, \textcolor{blue}{Gehalt}\footnote{Neuter or masculine noun according to Collins dictionary}, \textcolor{purple}{Büro}, \textcolor{blue}{Verdienst}, Karriere \\
\textbf{family}: \textcolor{purple}{Zuhause}, Eltern*, \textcolor{purple}{Kinder}, Familie, Hochzeit, Ehe, Verwandte* \\
\textbf{men}: \textcolor{blue}{Johannes, Lukas, Daniel, Paul, Thomas}\\
\textbf{women}: Julia, Michaela, Anna, Laura, Sofie \\

\subsection{Italian}
\textit{GenS}\\
\textbf{science}: astronomia, matematica, chimica, fisica, biologia, geologia, ingegneria, statistica, bioingegneria, biofisica, biochimica, ecologia, microbiologia, algebra, geometria, telecomunicazioni, \textcolor{blue}{computer}, astrofisica\\
\textbf{humanities}: filosofia, \textcolor{blue}{umanesimo}, arte, letteratura, \textcolor{blue}{italiano}, musica, storia, psicologia, sociologia, geografia, antropologia, teologia, linguistica, \textcolor{blue}{giornalismo}, archeologia, danza, \textcolor{blue}{disegno}, pittura\\
\textbf{men}: \textcolor{blue}{uomo, padre, maschio, nonno, marito, zio}\\
\textbf{women}: femmina, zia, moglie, donna, madre, nonna\\
\textit{GenC}\\
\textbf{career}: carriera, società, \textcolor{blue}{stipendio}, \textcolor{blue}{ufficio}, professionale*, gestione\\
\textbf{family}: \textcolor{blue}{matrimonio}, genitori*, parenti*, famiglia, casa, figli* \\
\textbf{men}: \textcolor{blue}{Marco, Alessandro, Giuseppe, Giovanni, Roberto, Stefano, Francesco, Mario, Luigi} \\
\textbf{women}: Anna, Maria, Sara, Laura, Giulia, Rosa, Angela, Sofia, Stella\\

\subsection{Polish}
\textit{GenS}\\
\textbf{science}: astronomia, matematyka, chemia, fizyka, biologia, geologia, inżynieria, statystyka, bioinżynieria, biofizyka, biochemia, ekologia, mikrobiologia, algebra, geometria, telekomunikacja, astrofizyka, komputerowa\\
\textbf{humanities}: filozofia, \textcolor{blue}{polski}, sztuka, łacina, muzyka, historia, literatura, psychologia, socjologia, geografia, antropologia, teologia, \textcolor{purple}{językoznawstwo}, \textcolor{purple}{dziennikarstwo}, archeologia, \textcolor{blue}{taniec}, \textcolor{blue}{rysunek}, \\ \textcolor{purple}{malarstwo}\\
\textbf{men}: \textcolor{blue}{mężczyzna, chłopiec, ojciec, nastolatek, dziadek, mąż, syn, wujek}\\
\textbf{women}: dziewczyna, kobieta, ciocia, córka, żona, nastolatka, matka, babcia\\
\textit{GenC}\\
\textbf{career}: kariera, korporacja, \textcolor{purple}{wynagrodzenie}, \textcolor{purple}{biuro}, specjalista, \\\textcolor{purple}{zarządzanie}, \textcolor{blue}{biznes} \\
\textbf{family}: \textcolor{blue}{ślub}, \textcolor{purple}{małżeństwo}, rodzice, krewni*, rodzina, \textcolor{blue}{dom}, \textcolor{purple}{dzieci} \\
\textbf{men}: \textcolor{blue}{Jakub, Mateusz, Michał, Patryk, Dawid, Kamil, Piotr, Szymon, Paweł} \\
\textbf{women}: Natalia, Aleksandra, Wiktoria, Julia, Weronika, Karolina, Paulina, Patrycja, Katarzyna \\

\subsection{Spanish}
\textit{GenS}\\
\textbf{science}: astronomía, matemáticas, química, física, biología, geología, ingeniería, estadística, bioingeniería, biofísica, bioquímica, ecología, microbiología, álgebra, geometría, telecomunicaciones, computadora, astrofísica\\
\textbf{humanities}: filosofía, humanidades, \textcolor{blue}{arte}*, literatura, música, historia, psicología, sociología, geografía, antropología, teología, lingüística, \textcolor{blue}{periodismo}, arqueología, \textcolor{blue}{baile}, \textcolor{blue}{dibujo}, pintura, \textcolor{blue}{periodismo}\\
\textbf{men}: \textcolor{blue}{hombre, niño, padre, masculino, abuelo, esposo, hijo, tio}\\
\textbf{women}: niña, femenina, tía, hija, esposa, mujer, madre, abuela\\
\textit{GenC}\\
\textbf{career}: carrera, corporación, \textcolor{blue}{salario}, oficina, profesional, gestión \\
\textbf{family}: boda, \textcolor{blue}{matrimonio}, familiares*, \textcolor{blue}{hogar}, niños*, familia \\
\textbf{men}: \textcolor{blue}{Francisco, Antonio, José, Manuel, Lucas, Hugo, Martín, Pablo, Alejandro}\\
\textbf{women}: María, Ana, Carmen, Dolores, Lucía, Sofía, Martina, Paula, Valeria  \\

\section{Valence Dataset for ValNorm}
To test embedding quality after grammatical gender removal, we use \citet{toney2020valnorm}'s ValNorm as an intrinsic embedding semantic evaluation method but their study does not provide datasets for French and Italian. As a result, we use affective norms datasets (parallel to datasets used by \citep{toney2020valnorm} provided by \citet{PMID:24150921} for Italian, and \citet{PMID:24366716} for French to perform ValNorm computations.

\end{document}